\documentclass[a4paper,11pt]{article}
\usepackage{multicol}
\usepackage{authblk}
\usepackage[utf8]{inputenc}

\usepackage{hyperref}
\usepackage{float}
\usepackage{amssymb}
\usepackage{amsmath}
\usepackage{array}

\begin{document}

\title{Emergent spin and clock variable in Bianchi type-I \\ quantum cosmology}
 
\author{Vishal\thanks{vishal.phiitg@iitg.ac.in}
\ and Malay K. Nandy\thanks{mknandy@iitg.ac.in}}

\affil{Department of Physics, Indian Institute of Technology Guwahati, Guwahati 781 039, India} 
  \date{February 20, 2024}

\maketitle           

\begin{abstract}
We consider the Bianchi type-I model of the universe in the Wheeler-DeWitt quantization scheme with the matter degree of freedom represented by a scalar field. As a consequence, the quantum mechanical equation of the universe is obtained in the minisuperspace consisting of the Misner variables and the scalar field. Employing Dirac factorization, we find that the volume parameter makes a suitable choice for the clock variable whereas the matter clock leads to various inconsistencies. We further find that the minisuperspace orbital angular momentum operator does not commute with the Hamiltonian in the Dirac-type equation. We therefore find the missing part of the angular momentum whereby the total angular momentum commutes with the Hamiltonian. We interpret this missing part as the spin of the quantum universe during its early stage of evolution. The emergence of the three-component spin vector is owing to the presence of anisotropy in the Bianchi type-I model of the universe which is absent in the quantization of isotropic models. 
\end{abstract}

\begin{multicols}{2}
\section{Introduction}
The observed large-scale distribution of matter in our universe is homogenous as well as isotropic while the universe is expanding at the same rate in every direction. The assumption of homogeneity and isotropy is further strengthened by the observed cosmic microwave background (CMB) radiation signifying homogeneity and isotropy even at the period of the last scattering. Thus a first step towards modeling  the universe involves the  Friedmann-Lemaitre-Robertson-Walker metric \cite{friedman1922krummung,lemaitre1927univers,robertson1935kinematics,walker1937milne}, $ds^2 = - dt^2  +a^2(t)\,(dx^2 +  dy^2 +  dz^2)$ (for a flat universe), where $a(t)$ is the scale factor and $t$ the cosmic time.

Although this isotropic model has been able to explain most of the macroscopic features of the universe, the anisotropies in the CMB (as well as in the large-scale matter distribution) call for anisotropic models. In particular, the anisotropies in the CMB power spectrum have been obtained by considering anisotropic fluctuations in the metric \cite{malik2009cosmological,mukhanov1992theory} so that the CMB anisotropies are explained as quantum mechanical perturbations in the inflationary era which serve to play the role of seeds in forming the anisotropic large-scale structure of the universe. The quantum nature of the primordial perturbations is an important feature that motivates the question of what might be the role of quantum gravity effects around the time of the creation of the universe.

Quantum gravitational corrections to the scalar perturbations were considered in the leading order in a universe undergoing de Sitter expansion \cite{kamenshchik2020quantum}. In addition, the 
quantum gravitational corrections to the scalar and tensor perturbations in the inflationary era were taken into account in the Wheeler-DeWitt (WDW) quantization scheme. It was found that the largest scales undergo an enhancement in the power spectrum \cite{brizuela2016quantum}.  

This motivates the question as to what might be the role of anisotropies in quantum gravity in the early stage of the universe. Anisotropies are expected to become prominent at early times and they need to be taken care of while constructing a theory of the universe.

Bianchi universes \cite{landau1971classical,ellis1999cosmological,ryan2015homogeneous,misner1972magic} are the simplest (classical) anisotropic models which are homogenous but do not respect spatial isotropy. In this work, we consider the Bianchi type-I model of the universe described by the metric
\begin{equation}
ds^2 = - N dt^2  +a^2(t) dx^2 + b^2(t) dy^2 + c^2(t) dz^2
\end{equation}
where the three unequal scale factors imply expansion of the universe with unequal rates in the three spatial directions.

We consider this anisotropic model in the minisuperspace representation of the Wheeler-DeWitt framework of quantum gravity \cite{WheelerQG,dewitt1967quantum} to study the behavior of the universe in the early stage of evolution. 

The Wheeler-DeWitt equation was obtained in the  ADM formalism of general relativity \cite{arnowitt1960canonical} which is based on decomposition of the spacetime into constant-time spatial hypersurfaces. This allows for the study of time evolution of the spacetime geometry as a Hamiltonian system. Quantization is achieved by promoting conjugate variables to operators in the Hamiltonian constraint to arrive at the WDW equation, $\hat{H} \Psi = 0$. This quantum gravity framework has been employed to investigate the initial quantum tunneling of the universe \cite{hartle1983wave, vilenkin1988quantum,vilenkin1982creation} as well as the cosmological singularity and CMB anisotropy in different models of the universe \cite{kiefer2019singularity,kiefer2012quantum}.

The Wheeler-DeWitt equation for FLRW and Bianchi minisuperspace models takes the form of the Klein-Gordon equation which suffers from the problem of negative probability density. To avoid this problem, Dirac factorization was employed to obtain a Dirac-type equation \cite{craig2010consistent,yamazaki2001dirac}. Upon Dirac factorization of the Wheeler-DeWitt equation for the vacuum Bianchi class A models, it was shown that the universe had a spin-like degree of freedom in the early quantal stage \cite{yamazaki2001dirac}.

In the present work, we incorporate a scalar field as a matter degree of freedom to the Bianchi type-I model in order to study the angular momentum of the early quantum state of the universe.

We find that the equation of continuity for the probability density and hermiticity of the Hamiltonian restrict the choice of the time variable. Upon choosing a legitimate time variable, we show that the orbital angular momentum, constructed from the minisuperspace variables, does not commute with the Hamiltonian. This signifies the existence of a missing part of the angular momentum. This missing part turns out to be originating from the spin degree of freedom. We demonstrate that the total angular momentum, orbital plus spin, does commute with the Hamiltonian. Thus the spin of the universe naturally emerges from the law of conservation of angular momentum. 
 
The outline of this paper is as follows. In section 2, we obtain the Wheeler-DeWitt equation for the Bianchi type-I universe in terms of Misner variables. In section 3, we factorize the WDW equation by Dirac factorization and obtain Schrodinger-type equations for different choices for the clock variable. It is shown that the volume parameter is suitable for the clock variable. In section 4, we calculate the commutation relation of the orbital angular momentum with the Hamiltonian and identify the missing part of the angular momentum as the spin of the quantum universe. In section 5, we summarise and discuss our results.
\section{Wheeler-DeWitt quantization}
For a spatially homogenous and anisotropically expanding universe, the Bianchi-I model is the simplest cosmological model. In order to obtain the Wheeler-DeWitt equation in this anisotropic model, we start with the Hamiltonian formalism of general relativity and write the Einstein-Hilbert action in ADM decomposed form,
\begin{equation}
S_G = \frac{1}{16 \pi G} \int dt \, d^3 x \; N \sqrt{h} \;({}^3\!R + K^{ij} K_{ij} - K^2),
\end{equation} 
where the  Ricci scalar of the four-dimensional spacetime is expressed in terms of the extrinsic curvature $K_{ij}$ and the Ricci scalar  ${}^3\!R$ on three-dimensional hypersurfaces \cite{poisson2002advanced}. Here $h$ is the determinant of the metric $h_{ij}$ defined on three-dimensional hypersurfaces and $K = K^{ij} h_{ij}$ is the trace of $K^{ij}$. For a Bianchi type-I universe, $K_{ij} = -\frac{\dot{h}_{ij}}{2N} = -\frac{1}{N} {\rm diag}(a \dot{a}, b \dot{b}, c \dot{c}) $ and ${}^3\!R = 0$. 

We use Misner variables  $(\alpha, \beta_+, \beta_-)$, defined by $a = e^{\alpha + \beta_{+} + \sqrt{3} \beta_{-}}$, $b = e^{\alpha + \beta_{+} - \sqrt{3} \beta_{-}}$ and $c = e^{\alpha -2 \beta_{+} }$ \cite{misner1969quantum,misner1970classical} so that $ e^\alpha = (abc)^{1/3}$ contains the volume information and the anisotropy parameters $\beta_+$ and $\beta_-$ determine the shape of the universe.

Integrating over the spatial part in the action generates the fiducial volume $V_0$ of the three-dimensional space and we choose units such that $3V_0/4 \pi G  = 1$. We consider a massless scalar field $\phi$ to represent the matter part in the universe. Consequently, the total action $S = S_G + S_m$, comprising of the gravitational and matter parts, is given by
\begin{equation}
S_G  =   \int dt \; \frac{1}{2N} \; e^{3 \alpha} (-\dot\alpha^{2}+ \dot\beta_{+}^{2} + \dot\beta_{-}^{2} )   
\end{equation} 
\begin{equation}
S_m =  \int dt \; \frac{1}{2N} e^{3 \alpha}\dot{\phi^{2}}  
\end{equation} 
Corresponding to the coordinate variables $(\alpha, \beta_+, \beta_-, \phi)$, we obtain the total Hamiltonian 
\begin{equation}
H =  \frac{N e^{- 3 \alpha}}{2} (-p_{\alpha}^{2} + p_{+}^{2} + p_{-}^{2}  + p_{\phi}^2)
\end{equation} 
in terms of the conjugate momenta $(p_\alpha, p_+, p_-, p_\phi)$. The classical Hamiltonian constraint reads
\begin{equation}
p_{\phi}^{2} -p_{\alpha}^{2} + p_{+}^{2} + p_{-}^{2}  = 0.
\end{equation}
Now we proceed to quantize the system by promoting the phase space variables to operators, ($ \hat{q}_a = q_a\; , \hat{p}_a = -i \hbar\frac{\partial }{\partial q_a } ) $ for $a = (\alpha, \beta_+,\beta_-, \phi)$. The resulting constraint now acts on the wavefunction $\Psi(\alpha, \beta_+, \beta_-, \phi)$ and yields the Wheeler-DeWitt equation
\begin{equation}
\left(\frac{\partial^2}{\partial \alpha^2} - \frac{\partial^2}{\partial \beta_{+}^2} - \frac{\partial^2}{\partial \beta_{-}^2}- \frac{\partial^2}{\partial \phi^2} \right) \; \Psi(\alpha, \beta_+, \beta_{-}, \phi) = 0.
\label{eq:Wd}
\end{equation}
This equation is of the Klein-Gordon type and we encounter the problem of negative probability density.
\section{Dirac factorization and clock variable}
For the resolution of the negative probability density following the above equation, we use Dirac factorization in this section. 
\subsection{Matter field $\phi$ as clock variable}
Upon choosing the matter field $\phi$ as the clock variable we rewrite the Wheeler-DeWitt equation \
(\ref{eq:Wd}) as
\begin{equation}
- \frac{\partial^2}{\partial \phi^2} \; \Psi(\vec{r}, \phi) = \Theta \; \Psi(\vec{r}, \phi)
\label{thetawd}
\end{equation} 
\begin{equation}
\Theta = (p_{\alpha}^2 - p_{+}^{2} - p_{-}^{2}) 
\end{equation}
where $\vec{r} = (\alpha, \beta_+, \beta_-)$.

To obtain a Dirac-type equation, we take the square root on both sides of the equation (\ref{thetawd}), giving
\begin{equation}
\pm i \frac{\partial}{\partial \phi} \Psi (\vec{r},\phi) = \sqrt{\Theta} \; \Psi(\vec{r}, \alpha) 
\label{timephi}
\end{equation}
with the square root of the Hamiltonian $\Theta$ expressed as
\begin{equation}
\sqrt{\Theta}  = (i \sigma_1 \hat{p}_+ + i \sigma_2 \hat{p}_- + \sigma_3 \hat{p}_\alpha) 
\label{thetasigma}
\end{equation}
where $\sigma_1,\sigma_2,\sigma_3$ are the Pauli matrices.

We note that the matter field $\phi$ plays the role of time in the Schrodinger type equation (\ref{timephi}). However, it is clearly seen that the corresponding Hamiltonian $\sqrt{\Theta}$ given by equation (\ref{thetasigma}) does not meet the condition of hermiticity.

To find the equation of continuity, we rewrite the above equation in the coordinate representation
\begin{equation}
\pm i \frac{\partial \Psi}{\partial \phi} - \sigma_1 \frac{\partial \Psi}{\partial \beta_{+}} - \sigma_2 \frac{\partial \Psi}{\partial \beta_{-}} + i \sigma_3 \frac{\partial \Psi}{\partial \alpha} = 0,
\label{eq:timephi1}
\end{equation}
and its Hermitian conjugate
\begin{equation}
\mp i \frac{\partial \Psi^{\dagger}}{\partial \phi} -  \frac{\partial \Psi^{\dagger}}{\partial \beta_{+}} \sigma_1 - \frac{\partial \Psi^{\dagger}}{\partial \beta_{-}} \sigma_2 - i \frac{\partial \Psi^{\dagger}}{\partial \alpha} \sigma_3 = 0.
\label{eq:timephi2}
\end{equation}
The above two equations lead to the equation of continuity: 
\begin{eqnarray}
i \frac{\partial}{\partial \phi} (\Psi^{\dagger} \Psi) + \left( \Psi^{\dagger} \sigma_1 \frac{\partial \Psi}{\partial \beta_+} - \frac{\partial \Psi^{\dagger}}{\partial \beta_+} \sigma_1 \Psi \right)\nonumber
\\+\left( \Psi^{\dagger} \sigma_2 \frac{\partial \Psi}{\partial \beta_-} - \frac{\partial \Psi^{\dagger}}{\partial \beta_-} \sigma_2 \Psi \right)  \nonumber \\
-i\left( \Psi^{\dagger} \sigma_3 \frac{\partial \Psi}{\partial \alpha} - \frac{\partial \Psi^{\dagger}}{\partial \alpha} \sigma_3 \Psi \right) = 0.
\end{eqnarray}
This equation deviates from the equation of continuity since the last line contains an imaginary contribution. This violates the law of conservation of probability.

To obtain the dispersion relation, we consider the plane wave solution in the form of a two-component spinor,
\begin{equation}
\Psi (r,\phi) = e^{i \vec{K}.\vec{r}} e^{i \omega \phi} 
\left(\begin{array}{cc}
\psi_1 \\
\psi_2 \\
\end{array} \right),
\end{equation}
where $ \vec{r} = (\beta_+ \; \hat{e}_+ + \beta_- \; \hat{e}_- + \alpha \; \hat{e}_\alpha) $ and $\vec{K} = (k_+ \; \hat{e}_+ + k_- \; \hat{e}_- + k_\alpha \; \hat{e}_\alpha) $. Employing this plane wave solution in Eq \ (\ref{timephi}), we get the dispersion relation 
\begin{equation}
\omega = \pm \sqrt{k_{\alpha}^2 - k_{+}^2 - k_{-}^2}.
\end{equation}
Thus, for modes with $ k_{+}^2 + k_{-}^2 >k_{\alpha}^2 $, this dispersion relation admits imaginary frequency solutions which is inconsistent with the plane wave solution. 

From the various inconsistencies encountered above, we conclude that the scalar field $\phi$ is not a good choice for the clock variable in a Bianchi type-I universe. 

\subsection{Volume parameter $\alpha$ as clock variable}
We shall now choose the volume parameter $\alpha$ as the clock variable. We rewrite the Wheeler-DeWitt equation (\ref{eq:Wd}) as
\begin{equation}
- \frac{\partial^2}{\partial \alpha^2} \; \Psi(\vec{r}, \alpha) = \Omega \; \Psi(\vec{r}, \alpha)
\end{equation}
\begin{equation}
\Omega = ( p_{+}^{2} + p_{-}^{2} + p_{\phi}^{2} ) 
\end{equation}
where $\vec{r} = (\beta_+, \beta_-, \phi)$.

Repeating the steps as before, we factorize this equation using Pauli matrices and obtain a Schrodinger-type equation
\begin{equation}
\pm i \frac{\partial}{\partial \alpha} \Psi (\vec{r},\alpha) = \sqrt{\Omega} \; \Psi(\vec{r}, \alpha),
\label{timealpha}
\end{equation}
\begin{equation}
\sqrt{\Omega} = (\sigma_1 \hat{p}_+ + \sigma_2 \hat{p}_- + \sigma_3 \hat{p}_\phi).   
\label{timealpha1}
\end{equation}
In the above Schrodinger equation, the operator $\sqrt{\Omega}$ acts as a Hamiltonian operator which is Hermitian since the momentum operators are themselves Hermitian and they commute with the Pauli matrices.

We can now proceed to obtain the continuity equation and check whether the law of conservation of probability is respected.

Rewriting equation (\ref{timealpha}) as 
\begin{equation}
\pm i \frac{\partial \Psi}{\partial \alpha} + i \sigma_1 \frac{\partial \Psi}{\partial \beta_{+}} + i \sigma_2 \frac{\partial \Psi}{\partial \beta_{-}} + i \sigma_3 \frac{\partial \Psi}{\partial \phi} = 0,
\label{eq:timealpha3}
\end{equation}
and its Hermitian conjugate being 
\begin{equation}
\mp i \frac{\partial \Psi^{\dagger}}{\partial \alpha} - i \frac{\partial \Psi^{\dagger}}{\partial \beta_{+}} \sigma_1  - i \frac{\partial \Psi^{\dagger}}{\partial \beta_{-}} \sigma_2 - i \frac{\partial \Psi^{\dagger}}{\partial \phi} \sigma_3  = 0,
\label{eq:timealpha4}
\end{equation}
we obtain the continuity equation:
\begin{eqnarray}
\frac{\partial}{\partial \alpha}(\Psi^{\dagger} \Psi) + \frac{\partial}{\partial \beta_+}(\Psi^{\dagger} \sigma_1 \Psi) \nonumber \\ + \frac{\partial}{\partial \beta_-}(\Psi^{\dagger} \sigma_2 \Psi) 
     +\frac{\partial}{\partial \phi}(\Psi^{\dagger} \sigma_3 \Psi) = 0.
\end{eqnarray}
This equation conforms with the continuity equation $\frac{\partial \rho}{\partial t} + \vec{\nabla} \cdot \vec{J} = 0$, with the identification $\rho = \Psi^{\dagger} \Psi$ and $\vec{J} = \Psi^{\dagger} \vec{\sigma} \Psi $. Thus this continuity equation meets the requirement of conservation of probability.

To analyze further, we take a plane wave solution for the two-component spinor $\Psi$ as
\begin{equation}
\Psi (\vec{r},\alpha) = e^{i \vec{K}.\vec{r}} e^{i \omega \alpha} 
\left(\begin{array}{cc}
\psi_1 \\
\psi_2 \\
\end{array} \right),
\end{equation}
with wave vector $\vec{K} = (k_+ \; \hat{e}_+ + k_- \; \hat{e}_- + k_\phi \; \hat{e}_\phi) $ and  $\vec{r} = (\beta_+ \; \hat{e}_+ + \beta_- \; \hat{e}_- + \phi \; \hat{e}_\phi) $. Employing this solution in equation (\ref{timealpha}) leads to the dispersion relation 
\begin{equation}
\omega = \pm \sqrt{k_{+}^2 + k_{-}^2 +  k_{\phi}^2},
\end{equation}
which supports solutions with only real-valued frequencies.

From the above consistencies, we thus conclude that the volume parameter $\alpha$ is a suitable clock parameter for the Bianchi type-I universe in the early stage of its quantum evolution.

\section{Angular momentum and spin}
We define the orbital angular momentum according to $\vec{L} = \vec{r} \times \vec{p}$ with ``position'' vector $ \vec{r} = (\beta_+ \; \hat{e}_+ + \beta_- \; \hat{e}_- + \phi \; \hat{e}_\phi) $ and the  ``momentum'' vector $\vec{p} = (p_+ \; \hat{e}_+ + p_- \; \hat{e}_- + p_\phi \; \hat{e}_\phi) $, so that 
\begin{eqnarray}
\vec{L} = (\beta_{-} p_{\phi} - p_{-} \phi) \; \hat{e}_+ - (\beta_{+} p_{\phi} - p_{+} \phi) \; \hat{e}_-  \nonumber \\  +(\beta_{+} p_{-} - p_{+} \beta_-) \; \hat{e}_\phi
\end{eqnarray}

With the three components of this ``orbital'' angular momentum operator, we obtain the following commutation relations with the Hamiltonian $\sqrt{\Omega}$,  
\begin{eqnarray}
&[L_+ ,\sqrt{\Omega}]& = i(\sigma_2 p_\phi- \sigma_3 p_-), \nonumber \\ 
&[L_- ,\sqrt{\Omega}]& = i(\sigma_3 p_+ - \sigma_1 p_\phi), \nonumber \\
&[L_\phi ,\sqrt{\Omega}]& = i(\sigma_1 p_- - \sigma_2 p_+).
\end{eqnarray}
Thus the ``orbital'' angular momentum $\vec{L}$ does not commute with the  Hamiltonian $\sqrt{\Omega}$. This implies that $\vec{L}$ is not a constant of motion.

In order to find the missing part of the angular momentum, we compute the commutation relations of the Pauli matrices with the Hamiltonian $\sqrt{\Omega}$ accounting for the fact that the Pauli matrices anticommute among themselves. We thus obtain 
\begin{eqnarray}
&[\sigma_1 , \sqrt{\Omega}]& = 2i (\sigma_3 p_- - \sigma_2 p_\phi), \nonumber \\
&[\sigma_2 , \sqrt{\Omega}]& = 2i (\sigma_1 p_\phi - \sigma_3 p_+),\nonumber \\
&[\sigma_3 , \sqrt{\Omega}]& = 2i (\sigma_2 p_+ - \sigma_1 p_-).
\end{eqnarray}
Combining the above two sets of commutation relations, we can easily see that the sum $ \vec{L} + \frac{1}{2} \vec{\sigma} $ commutes with the Hamiltonian $\sqrt{\Omega}$, that is
\begin{equation}
[\vec{L} + \frac{1}{2} \vec{\sigma}, \sqrt{\Omega}] = 0.
\end{equation}

Since the additional contribution $\frac{1}{2} \vec{\sigma}$ to the angular momentum originates from the Pauli matrices, it can be interpreted as the spin of the universe. Thus the Bianchi type-I universe behaves like a quantum particle with spin one-half in the early stage of its quantum evolution.
\section{Discussion and conclusion}
In this paper, we quantized the Bianchi type-I universe containing a matter field in the framework of the  Wheeler-DeWitt quantization scheme. This gives rise to a Klein-Gordon type equation in the Misner variables which naturally suffers from the problem of negative probability density.

To circumvent this problem, we first followed the usual choice of making the matter field as a clock variable and factorized the WDW equation to obtain a Dirac-type equation. However, this equation suffers from three difficulties: (a) the Hamiltonian in the Dirac-type equation does not meet the required condition of hermiticity, (b) the equation of continuity does not respect the law of conservation of probability, and (c) the plane wave solution allows for imaginary-valued frequencies. From these inconsistencies, we conclude that the scalar field does not make a good choice for the clock variable.

On the other hand, when we regarded the volume parameter $\alpha$ as the clock variable, all the above inconsistencies were found to be resolved: (a) the Hamiltonian in the Dirac-type equation satisfies the required condition of hermiticity, (b) the equation of continuity respects the law of conservation of probability, and (c) the plane wave solution allows for only real-valued frequencies. From these consistencies, we conclude that the volume parameter $\alpha$ is a good choice for the clock variable.

With the choice of volume parameter $\alpha$  as the clock variable, we obtained the commutation relation of the angular momentum operator $\vec{L}$ with the Hamiltonian $\sqrt{\Omega}$ of the Dirac type equation. However, we found that the orbital angular momentum  $\vec{L}$ is not a constant of motion as it does not commute with the Hamiltonian $\sqrt{\Omega}$. 

To find the missing angular momentum, we considered commutation relations of the Pauli matrices with the Hamiltonian $\sqrt{\Omega}$. We found that the combination $\vec{L} + \frac{1}{2} \vec{\sigma}$ commutes with Hamiltonian. The additional contribution $\frac{1}{2} \vec{\sigma}$ to the angular momentum can be interpreted as the spin of the universe since it originates from the Pauli matrices. We may say that the universe behaves like a quantum particle with spin one-half in the early stage of its quantum evolution.

The above conclusions rest on the Bianchi type-I model containing a matter field where the anisotropy of the universe is built into the model a priori. Since anisotropy is expected to be prominent during the early phase of the universe owing to the dominance of quantum mechanical fluctuations, these conclusions have the potential of realistic significance. It is clear that the emergence of the three-component spin vector is owing to the presence of anisotropy in the universe which is absent in the quantization of the isotropic models.

\section*{Acknowledgement}
Vishal is supported through a Research Fellowship from the Ministry of Education (MoE), Government of India.

\bibliographystyle{unsrt} 
\bibliographystyle{unsrt}

\end{multicols}
\end{document}